\documentclass[preprints,article,accept,pdftex,moreauthors]{Definitions/mdpi}

\firstpage{1} 
\pubvolume{1}
\issuenum{1}
\articlenumber{0}
\pubyear{2022}
\copyrightyear{2022}
\datereceived{} 
\dateaccepted{} 
\datepublished{} 
\hreflink{https://doi.org/10.48550/arXiv.\linebreak2203.17114} 
\doinum{10.48550/arXiv.2203.17114}
\pdfoutput=1

\usepackage{amssymb,amsfonts}
\usepackage{algorithmic}
\usepackage{textcomp}
\usepackage[nolist]{acronym}
\usepackage[normalem]{ulem}
\usepackage{comment}
\usepackage{color}
\usepackage{multicol,colortbl}
\usepackage{balance}
\usepackage{longtable}
\usepackage{relsize}
\usepackage{threeparttable}
\usepackage{makecell}
\usepackage{subcaption}
\usepackage{adjustbox}
\usepackage{array}
\newcolumntype{L}[1]{>{\raggedright\let\newline\\\arraybackslash\hspace{0pt}}m{#1}}
\newcolumntype{C}[1]{>{\centering\let\newline\\\arraybackslash\hspace{0pt}}m{#1}}
\newcolumntype{R}[1]{>{\raggedleft\let\newline\\\arraybackslash\hspace{0pt}}m{#1}}

\newcommand{\ste}[1]{\textcolor{red}{#1}}
\newcommand{\steremove}[1]{\ste{\sout{#1}}}

\newcommand{\felix}[1]{\textcolor[RGB]{217,95,2}{#1}}
\Title{A Methodology for Abstracting the Physical Layer of Direct V2X Communications Technologies}
\TitleCitation{A methodology for abstracting the physical layer of direct V2X communications technologies}

\Author{Zhuofei Wu $^{1,2}$\orcidB{}, Stefania Bartoletti $^{3}$\orcidA{},  Vincent Martinez$^{4}$\orcidC{}, and Alessandro Bazzi $^{2}$\orcidD{}}

\AuthorNames{Zhuofei Wu, Stefania Bartoletti, Vincent Martinez, and Alessandro Bazzi}

\AuthorCitation{Wu, Z.; Bartoletti, S.; Martinez, V.; Bazzi, A.}

\address{%
$^{1}$ \quad \textit{Harbin Engineering University}, Harbin, China; wzfhrb@hrbeu.edu.cn\\
$^{2}$ \quad \textit{WiLab, CNIT / DEI, University of Bologna}, Bologna, Italy; \{wu.zhuofei, alessandro.bazzi\}@unibo.it\\
$^{3}$ \quad \textit{CNIT/ DEE, University of Rome Tor Vergata}, Rome, Italy; stefania.bartoletti@cnit.it\\
$^{4}$ \quad \textit{NXP}, Toulouse, France; vincent.martinez@nxp.com
}

\begin{acronym} 
\acro{3GPP}{Third Generation Partnership Project}
\acro{2D}{two-dimensional}
\acro{2G}{second generation}
\acro{3G}{third generation}
\acro{4G}{fourth generation}
\acro{5G}{fifth generation}
\acro{6G}{sixth generation}
\acro{5GAA}{5G Automotive Association}
\acro{5GS}{5G system}
\acro{AF}{application function}
\acro{AIFS}{arbitration inter-frame space}
\acro{AMC}{adaptive modulation and coding}
\acro{AoI}{age of information}
\acro{AS}{application server}
\acro{AWGN}{additive white Gaussian noise}
\acro{BEP}{beacon error probability}
\acro{BM-SC}{Broadcast Multicast Service Center}
\acro{BP}{beacon periodicity}
\acro{B-CSA}{broadcast \ac{CSA}}
\acro{BSM}{basic safety message}
\acro{BSS}{basic service set}
\acro{BS}{base station}
\acro{C-ITS}{cooperative-intelligent transport systems}
\acro{C-NOMA}{Cooperative \ac{NOMA}}
\acro{C-V2X}{cellular-\ac{V2X}}
\acro{CA}{collision avoidance}
\acro{CAM}{cooperative awareness message}
\acro{CAV}{connected and autonomous vehicle}
\acro{CBR}{channel busy ratio}
\acro{CCDF}{complementary cumulative distribution function}
\acro{CDF}{cumulative distribution function}
\acro{CDMA}{code-division multiple access}
\acro{CoMP}{coordinated multi-point}
\acro{CN}{core network}
\acro{CP}{cyclic prefix}
\acro{CP-OFDM}{cyclic prefix orthogonal frequency-division multiplexing}
\acro{CPM}{collective perception message}
\acro{CR}{channel occupancy ratio}
\acro{CRDSA}{contention resolution diversity slotted ALOHA}
\acro{CSA}{coded-slotted ALOHA}
\acro{CSD}{cyclic shift diversity}
\acro{CSI}{channel state information}
\acro{CSIT}{channel state information at the transmitter}
\acro{CSIR}{channel state information at the receiver}
\acro{CSMA/CA}{carrier sense multiple access with collision avoidance}
\acro{CSSR}{candidate single-subframe resource}
\acro{D2D}{device-to-device}
\acro{DCC}{decentralized congestion control}
\acro{DCF}{distributed coordination function}
\acro{DCI}{downlink control
information}
\acro{DCM}{Dual carrier modulation}
\acro{DENM}{decentralized environmental notification message}
\acro{DMRS}{demodulation reference signal}
\acro{DRX}{discontinuous reception}
\acro{DSRC}{dedicated short range communication}
\acro{DOT}{Department of Transportation}
\acro{EC}{European Commission}
\acro{ECP}{extended cyclic prefix}
\acro{EDCA}{enhanced distributed coordination access}
\acro{EDGE}{Enhanced Data rates for GSM Evolution}
\acro{eMBMS}{evolved Multicast Broadcast Multimedia Service}
\acro{eMMB}{enhanced mobile broadband}
\acro{eNodeB}{evolved NodeB}
\acro{EPC}{Evolved Packet Core}
\acro{EPS}{Evolved Packet System}
\acro{ETSI}{European Telecommunications Standards Institute}
\acro{eV2X}{enhanced V2X}
\acro{FCC}{Federal Communications Commission}
\acro{FCD}{floating car data}
\acro{FD}{in-band full-duplex}
\acro{FDD}{frequency division duplex}
\acro{FDMA}{frequency division multiple access}
\acro{FEC}{forward error correction}
\acro{FR1}{frequency range 1}
\acro{FR2}{frequency range 2}
\acro{GNSS}{global navigation satellite system}
\acro{GPRS}{general packet radio service }
\acro{GPS}{global positioning system}
\acro{HARQ}{hybrid automatic repeat request}
\acro{HD}{half-duplex}
\acro{HSDPA}{High Speed Downlink Packet Access}
\acro{HSPA}{High Speed Packet Access}
\acro{IBE}{in-band emission}
\acro{ICI}{Inter-carrier interference}
\acro{IDMA}{interleave-division multiple access}
\acro{IM}{index modulation}
\acro{IP}{Internet Protocol}
\acro{IPG}{inter-packet gap}
\acro{IR}{incremental redundancy}
\acro{ISI}{inter-symbol interference}
\acro{ITS}{intelligent transport system}
\acro{i.i.d.}{independent identically distributed}
\acro{KPI}{key performance indicator}  
\acro{LDPC}{low-density parity-check}
\acro{LDS}{low-density spreading}
\acro{LIDAR}{light detection and ranging}
\acro{LOS}{line-of-sight}
\acro{LTE}{long term evolution}  
\acro{LTE-D2D}{\ac{LTE} with \ac{D2D} communications}  
\acro{LTE-V2X}{long-term-evolution-vehicle-to-anything} \acro{LTE-V2V}{\ac{LTE}-\ac{V2V}}
\acro{MAC}{medium access control}
\acro{MAE}{mean absolute error}
\acro{MBMS}{Multicast Broadcast Multimedia Service}
\acro{MBMS-GW}{MBMS Gateway}
\acro{MBSFN}{MBMS Single Frequency Network}
\acro{MCE}{Multi-cell Coordination Entity}
\acro{MEC}{mobile edge computing}
\acro{MCM}{maneuver coordination message}
\acro{MCS}{modulation and coding scheme}
\acro{MIMO}{multiple input multiple output}
\acro{ML}{machine learning}
\acro{MRD}{maximum reuse distance}
\acro{MUD}{multi-user detection}
\acro{NEF}{network exposure function}
\acro{NGV}{Next Generation V2X}
\acro{NHTSA}{National Highway Traffic Safety Administration}
\acro{NLOS}{non-line-of-sight}
\acro{NOMA}{non-orthogonal multiple access}
\acro{NOMA-MCD}{\ac{NOMA}-mixed centralized/distributed}
\acro{NR}{new radio}
\acro{OBU}{on-board unit}
\acro{OCB}{outside of the context of a basic service set}
\acro{OEM}{Original equipment manufacturer}
\acro{OFDM}{orthogonal frequency-division multiplexing}
\acro{OFDMA}{orthogonal frequency-division multiple access}
\acro{OMA}{orthogonal multiple access}
\acro{OTFS}{orthogonal time frequency space}
\acro{pdf}{probability density function}
\acro{PAPR}{peak to average power ratio}
\acro{PCF}{policy control function}
\acro{PCM}{platoon control message}
\acro{PD}{packet delay}
\acro{PEP}{pairwise error probability} 
\acro{PER}{packet error rate}
\acro{PIAT}{packet inter-arrival time}
\acro{PRB}{physical resource block}
\acro{PSCCH}{Physical Sidelink Control Channel}
\acro{PSFCH}{Physical Sidelink Feedback channel}
\acro{PSSCH}{Physical Sidelink Shared channel}
\acro{PHY}{physical}
\acro{PL}{path loss}
\acro{PLMN}{Public Land Mobile Network}
\acro{PPDU}{Protocol Packet Data Unit}
\acro{PPPP}{ProSe Per-Packet Priority}
\acro{PPPR}{ProSe Per-Packet Reliability}
\acro{ProSe}{Proximity-based Services}
\acro{PRR}{packet reception ratio}
\acro{QAM}{quadrature amplitude modulation}
\acro{QC-LDPC}{quasi-cyclic \ac{LDPC}}
\acro{QoS}{quality of service}
\acro{QPSK}{quadrature phase shift keying}
\acro{RAN}{Radio Access Network}
\acro{RAT}{radio access technology}
\acro{RE}{resource element}
\acro{RF}{radio frequency}
\acro{RIS}{reconfigurable intelligent  surface}
\acro{RL}{reinforcement learning}
\acro{RR}{radio resource}
\acro{RRC}{radio resource control}
\acro{RRI}{resource reservation interval}
\acro{RSRP}{reference signal received power}
\acro{RSSI}{Received Signal Strength Indicator}
\acro{RSU}{road side unit} 
\acro{SAE}{Society of Automotive Engineers}
\acro{SAI}{Service Area Identifier}
\acro{SB-SPS}{sensing-based semi-persistent scheduling}
\acro{SC-FDMA}{single carrier frequency division multiple access}
\acro{SC-PTM}{Single Cell Point To Multipoint}
\acro{SCS}{subcarrier spacing}
\acro{SCMA}{sparse-code multiple access}
\acro{SI}{self-interference}
\acro{SIC}{successive interference cancellation}
\acro{S-UE}{scheduling UE}
\acro{SCI}{sidelink control information}
\acro{SDN}{Software-defined networking}
\acro{SFFT}{symplectic finite Fourier transform}
\acro{SHINE}{simulation platform for heterogeneous interworking networks}
\acro{SINR}{signal-to-interference-plus-noise ratio}
\acro{SNR}{signal-to-noise ratio}
\acro{SRS}{sounding reference signal}
\acro{STBC}{space time block codes}
\acro{SV}{smart vehicle}
\acro{TB}{transport block}
\acro{TBC}{time before change}
\acro{TBE}{time before evaluation}
\acro{TDD}{time-division duplex}
\acro{TDMA}{time-division multiple access}
\acro{TR}{technical report}
\acro{TS}{technical specification}
\acro{TTI}{transmission time interval}
\acro{UAV}{unmanned aerial vehicle}
\acro{UD}{Update delay}
\acro{UMTS}{universal mobile telecommunications system}
\acro{URLLC}{ultra-reliable and ultra-low latency communications}
\acro{USIM}{Universal Subscriber Identity Module}
\acro{UTDOA}{uplink time difference of arrival}
\acro{V2C}{vehicle-to-cellular} 
\acro{V2N}{vehicle-to-network}
\acro{V2I}{vehicle-to-infrastructure} 
\acro{V2P}{vehicle-to-pedestrian}
\acro{V2R}{vehicle-to-roadside}
\acro{V2V}{vehicle-to-vehicle} 
\acro{V2X}{vehicle-to-everything} 
\acro{VAM}{VRU awareness message}
\acro{VRU}{vulnerable road user}
\acro{VUE}{vehicular user equipment}
\acro{WAVE}{wireless access in vehicular environment}
\acro{WBSP}{wireless blind spot probability}
\end{acronym}

\abstract{Recent advancements in V2X communications have greatly increased the flexibility of the physical and \ac{MAC} layers. This increases the complexity when investigating the system from a network perspective to evaluate the performance of the supported applications. Such flexibility needs in fact to be taken into account through a cross-layer approach, which might lead to challenging evaluation processes. As an accurate simulation of the signals appears unfeasible, a typical solution is to rely on simple models for incorporating the physical layer of the supported technologies, based on off-line measurements or accurate link-level simulations. Such data is however limited to a subset of possible configurations and extending them to others is costly when not even impossible. The goal of this paper is to develop a new approach for modelling the physical layer of \ac{V2X} communications that can be extended to a wide range of configurations without leading to extensive measurement or simulation campaign at the link layer. In particular, given a scenario and starting from results in terms of \ac{PER} vs. \ac{SINR} related to a subset of possible configurations, we derive one parameter, called implementation loss, that is then used to evaluate the network performance under any configuration in the same scenario. The proposed methodology, leading to a good trade-off among complexity, generality, and accuracy of the performance evaluation process, has been validated through extensive simulations with both IEEE~802.11p and LTE-V2X sidelink technologies in various scenarios. 
}


\begin{document}

\acresetall

\section{Introduction}
Vehicle-to-everything (V2X)\acused{V2X} connectivity allows vehicles to communicate with one another and with other road elements to share local views and intentions, discover surroundings, and  coordinate driving maneuvers, improving the safety and efficiency of our transportation systems  \cite{bazzi2021design,9497103,SOTO2022100428}. 
Focusing on direct communications, two families of standards have been defined for \ac{V2X} connectivity, i.e., the one based on IEEE~802.11p, which is expected to be shortly amended by the IEEE~802.11bd, and the other based on the sidelink technologies designed by the 3GPP for \ac{V2X}, which means today \ac{LTE}\acused{LTE-V2X} and 5G, and might become 6G in the next decade. 

The new developments in V2X standardization have enabled greater flexibility at both the \ac{PHY} and \ac{MAC} layers. This calls for a cross-layer performance analysis, where the main \ac{PHY} and \ac{MAC} layer parameters and procedures, as well as the interplay between them, should be considered under a variety of different scenarios and settings. One of the main issues when investigating V2X over multiple layers is how to abstract the PHY layer in a sufficiently accurate way without overly impacting on the computational complexity. An accurate simulation of the PHY layer would in fact require, in principle, a bit-by-bit generation per each transmitter and every receiver, a conversion to electromagnetic signals, and the propagation through a multi-path variable channel; clearly, this kind of approach is inconvenient when dealing with network-level simulator, which usually take into account large vehicle densities and different levels of mobility. Several approaches have been proposed to reduce the burden of PHY layer simulation in the evaluation of various communication technologies (i.e., not only V2X), which are tailored to the specific communications technology under investigation \cite{papanastasiou2010bridging,ben2009impact,BAZZI201247}.

Recently, with the rapid advances of V2X communications, the abstraction of PHY layer for V2X technologies  has attracted renovated attention \cite{Fallah2019,anwar_2018,Anwar_2021}. 
In this context, the channel normally varies quickly enough  that the small-scale fading observed by different transmissions in the time domain can be assumed as uncorrelated; additionally, transmissions are normally in broadcast mode, leading to a high number of links to be evaluated, which increase with the square of the density of the nodes. For these reasons, the correct reception is normally simulated by relying on the average \ac{PER} as a function of the \ac{SINR}, see e.g. Fig.~\ref{fig:PERmeasured}. The  PER vs. SINR curve can be derived through link-level simulations or experimentally. Then, the \ac{SINR} at each transmission is calculated by taking into account the impact of path-loss and large-scale fading; given the \ac{SINR}, the fate of the transmission is statistically determined based on the corresponding  \ac{PER}. 

Despite this is a widely adopted solution, the link-level simulations or experimentations used for generating these curves are usually computationally intensive or operationally unfeasible. Moreover, a different \ac{SINR} vs. \ac{PER} curve has to be generated for each scenario or system setting, i.e. any technology, packet size, \ac{MCS}. Note that there exist a plethora of configurations for \ac{V2X} communications, as the packet size is variable \cite{MarBer:18} and the \ac{MCS} might be adapted to channel conditions \cite{7906621,www:QualcommMCS}. Alternatively to the use of experimental PER vs. SINR curves, in \cite{Fallah2019}, the power of the signal received from each source is derived taking into account path-loss, shadowing, and fading; then, the \ac{SINR} is calculated and compared to a given threshold to assess the correctness of the reception. Although the abstraction makes the simulation fast,  
the threshold is calculated ad-hoc for a specific scenario and configuration. 

In \cite{anwar_2018,Anwar_2021}, a more accurate modelling of the propagation is proposed, where the \ac{SINR} is calculated on a per-subcarrier basis and an effective SINR is then derived through a parameter that needs optimization. In this case, a first drawback is that the mentioned parameter needs optimization for each adopted modulation. More importantly, a second drawback is that the channel transfer function needs to be calculated for each transmitter and receiver link, which severely impacts on the speed of the simulation.

In this paper, we propose a methodology for the \ac{PHY} abstraction of \ac{V2X} communications, which extends the PHY-level results available for few configurations; it can be used at the network level for mathematical models or simulations, in the latter case with very reduced impact on the processing and memory consumption. Specifically, we present a methodology to derive a parametric model, with a single parameter called implementation loss that depends on the operating scenario. We first approximate the PER vs. SINR curves with step functions, i.e. the packet is correctly received if the SINR is above a given threshold and discarded if it is below. Such approximation is shown to be sufficiently accurate for the most relevant configurations of traffic densities and technologies. Then,  the SINR threshold associated to the specific configuration under test is derived by leveraging the notion of effective throughput and the calculated implementation loss. 

The proposed methodology is validated by using it in network-level simulations. As a benchmark, the same evaluations are also carried out by relying on the PER vs. SINR curves obtained through link-level simulations, which are accurate yet computationally intensive and limited to a few configurations. Results show that the model resulting from the proposed methodology leads to an accurate evaluation at the network level of direct \ac{V2X} communications technologies, with a negligible impact on the processing speed and being able to cover a high number of relevant cases without the need of additional and heavy campaigns of measurements or link-level simulations.

\section{V2X Technologies}
The main families of technologies for direct V2X communications are currently those based on IEEE 802.11p and those under the umbrella of \ac{C-V2X} and denoted as sidelink. The two families rely on \ac{OFDM} at the \ac{PHY} layer and differ for the access mechanisms at the \ac{MAC} layer \cite{VUKADINOVIC201817}. In this section, we recall the mechanisms at the \ac{MAC} layer for both of them and define the transmission time for a generic payload of $P_\text{b}$ bytes. The duration of the generic transmission is required for the calculation of the effective throughput in Section~\ref{sec:model}.


\subsection{IEEE 802.11p}


IEEE 802.11p is an approved amendment to the IEEE 802.11 standard for the physical and \ac{MAC} layer of vehicular communications. In the PHY layer, IEEE 802.11p operates in the 5.9~GHz ITS band and uses \ac{OFDM} with 10 MHz bandwidth. Each \ac{OFDM} symbol includes 52 subcarriers with subcarrier spacing of 156.25~kHz (4 of them used as pilot), and lasts 8 $\mu$s. There are 8 possible combinations of \ac{MCS}, with the modulation ranging from BPSK to 64-QAM and the encoding implemented through a convolutional code with rate 1/2, possibly punctured to reach 2/3 or 3/4. The signal transmitted at the PHY layer consists of preamble field (32 $\mu$s), signal field (8 $\mu$s), and data field (variable time). More details can  be found in \cite{abdelgader2014physical}. 

The MAC algorithm deployed by IEEE 802.11p is called enhanced distributed coordination access (EDCA). It is based on the basic distributed coordination function (DCF) but adds QoS attributes. DCF is a carrier sense multiple access with collision avoidance (CSMA/CA) algorithm. In CSMA/CA, a node listens to the channel before transmission and if the channel is perceived as idle for a predetermined time interval the node starts to transmit. If the channel becomes occupied during such interval, the node performs a backoff procedure, i.e. the node defers its access according to a randomized time period. In IEEE 802.11p, the predetermined listening period is called \ac{AIFS} \cite{harigovindan2012ensuring}. 
Therefore, we can calculate the time required to transmit a packet with a given payload $P_\text{b}$ on the
wireless medium as \cite{anwar2019physical}
\begin{align}
 \label{eq:time11p}
    T_\text{tx}^{(11\text{p})} = T_\text{AIFS} + T_\text{pre} + T_\text{sym} × n_\text{sym}
\end{align}
where $T_\text{AIFS}$ is the duration of the AIFS, $T_\text{pre}$ is the preamble duration ($40\;\mu$s, including the preamble field and the signal field),
$T_\text{sym}$ is the \ac{OFDM} symbol duration ($8\;\mu$s),
and $n_\text{sym} = \lceil 8 P_b/n_\text{bpS}\rceil$ denotes the number of \ac{OFDM} symbols required to transmit a certain
payload (including MAC header, service, and tails bits), and $n_\text{bpS}$ is the number of data bits per \ac{OFDM} symbol \cite{etsi302663}.

\subsection{C-V2X sidelink}
At the lower layers,  sidelink numerology and building blocks of \ac{C-V2X} are based on the uplink specifications, which are \ac{SC-FDMA} in \ac{LTE-V2X} and \ac{CP-OFDM} in 5G-V2X. \ac{LTE-V2X} operates in 10~MHz or 20~MHz channels, whereas 5G-V2X can occupy up to 100~MHz when used in bands below 6~GHz (namely, sub 6~GHz). The resources are 
based on a time-frequency matrix structure, where the time domain is divided into transmission time interval (TTI), of 1~ms duration in LTE-V2X and of either 0.25~ms, 0.5~ms, or 1~ms in 5G-V2X (sub 6~GHz).

In the frequency domain, radio resources are organized in \acp{RE}, which aggregate into \acp{PRB}, in turn realizing the subchannels. 
Each RE is a subcarrier (spaced by 15~kHz in LTE and 15, 30, or 60~kHz in 5G) over an \ac{OFDM} symbol. Each PRB is composed of 12 consecutive subcarriers in frequency domain with the same subcarrier spacing (SCS). The sub-channels are composed of a certain number of PRBs. 
As the SCS changes, the bandwidth of a PRB varies accordingly. As a result, the number of PRBs and subchannels within a fixed channel bandwidth depends on the SCS.

A packet is normally transmitted on one or more subchannels within one TTI, which lasts 1~ms in LTE-V2X and either 0.25, 0.5, or 1 ms in 5G-V2X, depending on the SCS. In principle the transmission can be split over more than one TTI if the packet size and adopted MCS require more subchannels than those that are available. Therefore, we can calculate the time required to transmit a packet as
 \begin{align}
 \label{eq:timeCV2X}
     T_\text{tx}^{(\text{C-V2X})}= T_\text{TTI} \Big\lceil \frac{n_\text{PRB-pkt}}{n_\text{PRB-TTI}}\Big\rceil= T_\text{TTI} \cdot n_\text{TTI}
 \end{align}
 where $T_\text{TTI}$ is the TTI duration, $n_\text{PRB-pkt}$ is the number of \acp{PRB} necessary for one packet transmission (which depends on $P_\text{b}$ and the adopted \ac{MCS} \cite{bazzi2017performance}), $n_\text{PRB-TTI}$ is the number of PRBs in a TTI \cite{3GPP36331}, and $n_\text{TTI}=\Big\lceil \frac{n_\text{PRB-pkt}}{n_\text{PRB-TTI}}\Big\rceil$ is the number of TTIs needed for transmitting the packet.  In most of the cases, the transmission lasts a single TTI, thus $n_\text{TTI}=1$ and $T_\text{tx}^{(\text{C-V2X})}= T_\text{TTI}$.

\section{Physical Layer Abstraction Methodology}\label{sec:model}

In this section, we propose a general methodology to leverage a set of available curves under specific settings to obtain a more general PHY layer abstraction to be used in the network level simulations of \ac{V2X} communications. We first present the main idea and assumptions, and then detail the methodology.

As a starting point, we approximate the PER vs. SINR curves using step functions, where each step corresponds to a certain threshold of SINR $\gamma_\text{th}$. Note that the use of a step function to approximate the PER vs SINR curve is a common choice as it considers a packet as correctly received if the \ac{SINR} is above the threshold $\gamma_\text{th}$  \cite{sepulcre2022analytical, campolo2019NR}. Results shown in Section~\ref{sec:per_value} demonstrate that the impact of such an approximation is very limited when focusing on network-level simulations. The scope of the proposed approach is to derive SINR thresholds  $\hat{\gamma}_\text{th}$ that can be easily calculated for any settings and technologies, including those for which the PER vs SINR curves are not available.

\begin{figure}
\centering
\includegraphics[keepaspectratio,width=0.8\columnwidth]{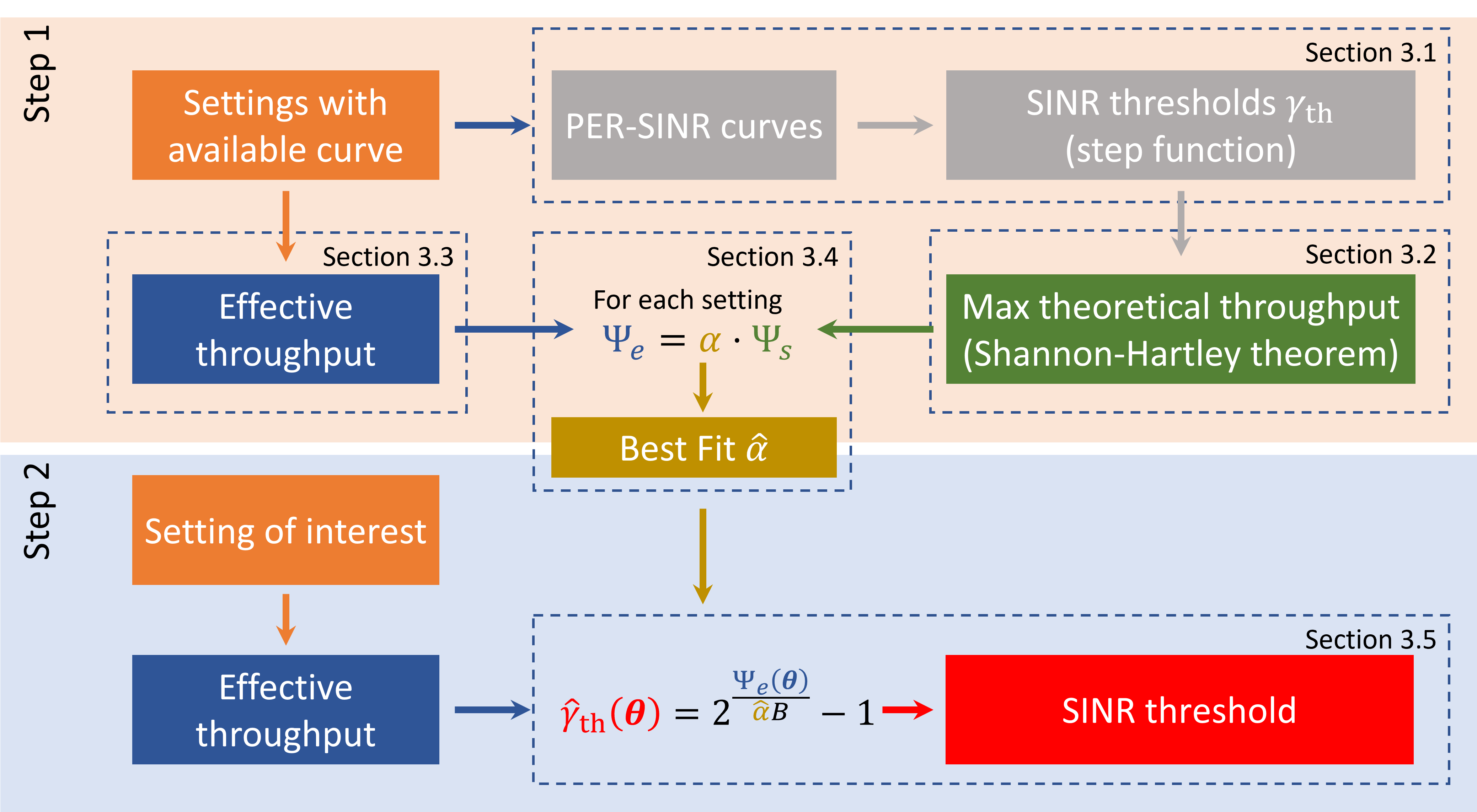}
\caption{Illustration of the two main steps defining the proposed methodology for a given scenario. The first step is deriving a best fit $\hat{\alpha}$ to approximate the effective throughput based on the PER vs. SINR curves that are available (Section~\ref{sec:attenuation_factor}). The second step is to derive the SINR threshold for the settings of interest using the calculated effective throughput and the derived $\hat{\alpha}$.}
\label{fig:summary}
\end{figure}

The methodology is illustrated in Fig.~\ref{fig:summary}: we start from a given scenario, for which PER vs. SINR curves are available for a subset of configurations, i.e., for some technologies, \acp{MCS}, and packet sizes; then, we calculate a parameter called \textit{implementation loss} and denoted as $\hat{\alpha}$, which is  used to derive the SINR thresholds  $\hat{\gamma}_\text{th}$ for any other configurations in that scenario. In our work, curves are available for an highway scenario in \ac{LOS} and \ac{NLOS} conditions and for an urban scenario in \ac{LOS} and \ac{NLOS} conditions, in all cases for a set of settings of both IEEE 802.11p and LTE-V2X sidelink.



In the first step, given a specific scenario and the system settings for a single PER vs. SINR curve, we start obtaining the optimal SINR threshold as described in Section~\ref{sec:approximation}. Based on this, we use the Shannon-Hartley theorem to calculate the maximum throughput in an \ac{AWGN} channel $\Psi_\text{s}$ for that SINR value (Section~\ref{subsec:maxThr}). At the same time, we calculate the \textit{effective throughput} $\Psi_\text{e}$ of the given technology and settings as detailed in Section~\ref{subsec:effThr}. Then, we assume that the effective throughput $\Psi_\text{e}$ can be approximated as an attenuated form of the maximum throughput $\Psi_\text{s}$, which is a function of the SINR threshold, i.e.,
\begin{align}
\label{eq:Shannonapprox}
    \Psi_\text{e}( \boldsymbol \theta)\simeq\alpha  \Psi_\text{s}(\gamma_\text{th}( \boldsymbol \theta))
\end{align}
where $\alpha$ is the implementation loss for that specific scenario and configuration. Note that an equation similar to \eqref{eq:Shannonapprox} to approximate the effective throughput starting from a given SINR is often used, an example being \cite{3GPP_36_942}. The first step is concluded, as detailed in Section~\ref{sec:attenuation_factor}, by calculating a single value for the implementation loss, denoted as $\hat{\alpha}$, which is the value that best approximates those obtained from the available curves.



In the second step, for any technology and system settings of interest, the value of $\Psi_\text{e}$ is calculated and then used together with the implementation loss $\hat{\alpha}$ of the first step to derive the SINR threshold $\hat{\gamma}_\text{th}$, as detailed in Section~\ref{subsec:genericSINRthreshold}.

\subsection{PER vs. SINR Curve Approximation}
\label{sec:approximation}
\begin{figure}
\centering
\includegraphics[keepaspectratio,width=0.8\columnwidth]{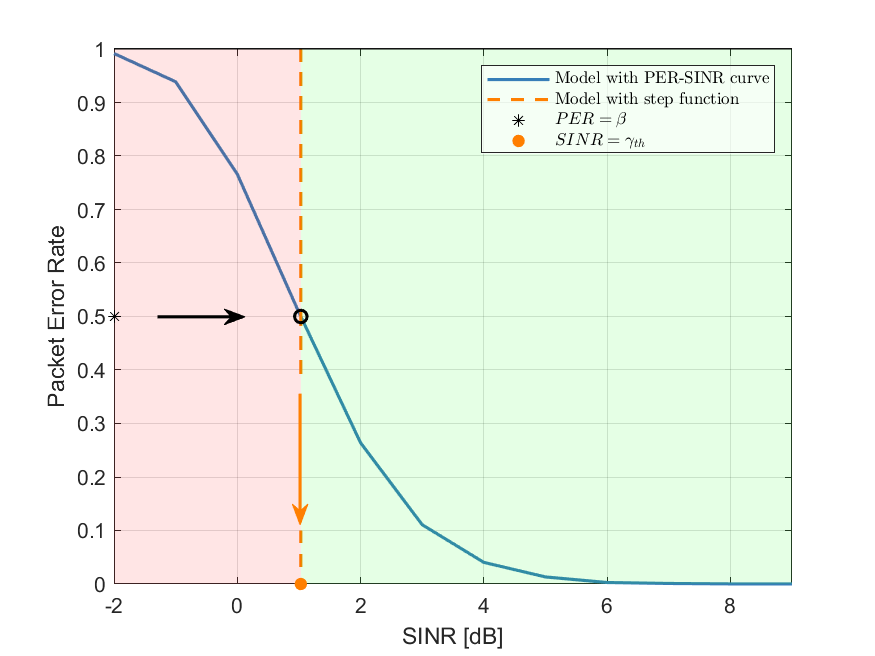}
\caption{From the PER vs. SINR curve (solid blue) to the approximating step function (dashed orange). The asterisk is the target PER value $\beta$ and the orange point indicates the corresponding SINR threshold $\gamma_{th}$. Adopting the step function, the packet is assumed successfully received  if the SINR is higher than the threshold (light green part), and not received otherwise (light red part).}
\label{fig:example}
\end{figure}

Fig.~\ref{fig:example} illustrates the method for the derivation of $\gamma_{th}$ from the PER vs. SINR curve. The blue solid line is a given PER vs. SINR curve corresponding to a specific scenario and certain system settings. We obtain $\gamma_{th}$ as the SINR value that corresponds on the curve to a certain PER value $\beta$ (the asterisk). It follows that instead of the original PER vs. SINR curve, we now have a step function (represented through  an orange dashed line in Fig.~\ref{fig:example}). In order to determine which value of $\beta$ to use, later called $\hat{\beta}$, the \ac{MAE} is used, calculated through the use of network level simulations. In particular, the MAE is calculated looking at the \ac{PRR} (i.e., the percentage of the packets correctly received at a given distance)  varying the source-destination distance as:
\begin{align}
    MAE = \frac{1}{n} \cdot \sum_{i=1}^{n} \left|  \text{PRR}_i^{(sf)}(\beta)-\text{PRR}_i^{(curve)} \right| 
\end{align}
where $\text{PRR}_i^{(sf)}(\beta)$ and $\text{PRR}_i^{(curve)}$ are the $i$-th PRR value point in the PRR vs. distance curves (e.g., see Fig.~\ref{fig:evaluate_beta}). By minimizing the MAE, the best value $\hat{\beta}$ is obtained.

\subsection{Maximum Throughput $\Psi_\text{s}$}\label{subsec:maxThr}

From the SINR threshold $\gamma_{th}$, the channel capacity as defined by the Shannon-Hartley theorem, i.e. the maximum theoretical throughput $\Psi_\text{s}$ that can be achieved over an \ac{AWGN} channel for a given \ac{SINR}, is calculated as 
\begin{align}
\label{eq:Shannon}
    \Psi_\text{s}(\gamma_\text{th})=  B \log_2(1+\gamma_\text{th})
\end{align}
where $B$ is the bandwidth of the channel and $\gamma_\text{th}$ is the \ac{SINR} threshold. 

\subsection{Effective Throughput $\Psi_\text{e}$}\label{subsec:effThr}

The effective throughput is defined as the maximum net throughput for the given configuration \cite{anwar2019physical}. In particular, given the packet size $P_\text{b}$ and the \ac{MCS}, the effective throughput is calculated as the ratio between the number of data bits and the time required for the transmission, which means for IEEE 802.11p and C-V2X that it can be calculated using \eqref{eq:time11p} and \eqref{eq:timeCV2X} as:
\begin{align}
   \Psi_\text{e}^{(11\text{p})}(\boldsymbol \theta^{(11\text{p})})  &= \frac{8P_\text{b}}{T_\text{tx}^{(11\text{p})}}\\
   \Psi_\text{e}^\text{(C-V2X)}(\boldsymbol \theta^{(\text{C-V2X})}) &=\frac{8P_\text{b}}{T_\text{tx}^{(\text{C-V2X})}} \cdot \frac{n_\text{subch} \cdot n_\text{PRB-subch}}{n_\text{PRB-pkt}}\,
\end{align}
 where $\boldsymbol \theta^{(11\text{p})}$ and $\boldsymbol \theta^{(\text{C-V2X})}$ represent the generic system setting vectors for IEEE 802.11p and C-V2X, respectively, i.e.:
 \begin{align}
    \boldsymbol \theta^{(11\text{p})}&=[T_\text{pre}, T_\text{AIFS}, T_\text{sym}, n_\text{sym}]\\
    \boldsymbol \theta^{(\text{C-V2X})}&=[n_\text{subch}, n_\text{PRB-subch}, T_\text{TTI}, n_\text{PRB-pkt}]
 \end{align}
with $n_\text{subch}$ being the number of subchannels and $n_\text{PRB-subch}$ the subchannel size, expressed as number of \acp{PRB}. Please note that the number of PRBs in a TTI can be written as a function of the number of subchannels and PRBs per subchannel as $n_\text{PRB-TTI} = n_\text{subch} \cdot n_\text{PRB-subch}$.

\subsection{Best Fit Implementation Loss $\hat{\alpha}$}
\label{sec:attenuation_factor}
Per each PER vs. SINR curve, the operations detailed in Sections~\ref{sec:approximation}, \ref{subsec:maxThr}, and~\ref{subsec:effThr} can be used to calculate the effective throughput and the SINR threshold, which in principle allow to obtain the implementation loss $\alpha$ using \eqref{eq:Shannonapprox}. However, in the general case, only the effective throughput can be calculated and both the SINR threshold and the implementation loss $\alpha$ are unknown. In order to relate the effective throughput with the SINR threshold for any possible configuration, the best fit $\hat{\alpha}$ which best approximates the value of $\alpha$ in the known cases is derived. 

Specifically, assume that there are $N$ available PER vs. SINR curves within a specific scenario. Each curve corresponds to specific parameter settings, i.e. $\{\boldsymbol \theta_i|i=1,2,\ldots,N\}$, where $\boldsymbol \theta_i$ represents a vector that includes the \ac{PHY} and \ac{MAC} parameters for the $i$-th settings.
In order to estimate the parameter $\hat{\alpha}$, a least-square approach is considered over the set of available curves, i.e.  
\begin{align}
\label{eq:estimatedalpha}
    \hat{\alpha}= \arg\min_{\alpha}
    \sum_{i=1}^{N} \left[ \Psi_\text{e}(\boldsymbol \theta_i) - \alpha \Psi_\text{s}(\gamma_\text{th}( \boldsymbol \theta_i))\right]^2
\end{align}

\subsection{SINR Threshold for the Generic Settings}
\label{subsec:genericSINRthreshold}

Once obtained the value $\hat{\alpha}$ for the given scenario, as explained in Section~\ref{sec:attenuation_factor}, it can be used for any parameter setting $\boldsymbol \theta$ beyond those for which a PER vs. SINR curve is available, e.g. for any MCS and for any packet size. The \ac{SINR} threshold correpsonding to the generic $\boldsymbol \theta$ is in fact obtained by combining \eqref{eq:Shannonapprox} and \eqref{eq:Shannon} as
\begin{align}
\label{eq:get_threshold}
    \hat{\gamma}_\text{th}( \boldsymbol \theta)= 2^{\frac{\Psi_\text{e}( \boldsymbol \theta)}{\hat{\alpha} B}}-1
\end{align}

\begin{table}
\caption{Main simulation parameters and settings
\label{Tab:Settings}}
\centering
\adjustbox{width=\textwidth}{
    \begin{tabular}{ll}
    \hline \hline
    \emph{\textbf{Scenario}} & \\
    Road layout & Highway, 3+3 or 6+6 lanes, 4 m width \\
    (Density, Average speed) [vehicles/km, km/h] & (100, 96) \& (400, 56)\\
    \hline
    \emph{\textbf{Power and propagation}} & \\
    Channels and Bandwidth & ITS 10 MHz bands at 5.9 GHz \\
    Transmission power density & 13~dBm/MHz \\
    Antenna gain (tx and rx) \& Noise figure  & 3 dBi \& 6~dB\\
    Propagation model & WINNER+, Scenario B1 \\
    Shadowing & Variance 3 dB, decorr. dist. 25~m \\
    \hline
    \emph{\textbf{Data traffic}}& \\
    Packet size \& generation rule  & $P_\text{b}=350$~bytes \&  Following the rules in \cite{etsi201904} \\ 
    \hline
    \emph{\textbf{IEEE 802.11p settings}} & \\
    MCS & 2 (QPSK, CR$=0.5$)\\
    Maximum contention window & 15\\
    Arbitration inter-frame space & 110 $\mu s$\\
    Sensing threshold for known \& unknown signals & -85 dBm \& -65 dBm\\\hline
    \emph{\textbf{Sidelink LTE-V2X settings}} & \\
    MCS & 7 (QPSK, CR$\approx0.5$) \\
    Number and size of subchannels $n_\text{subch}$ \& $n_\text{PRB-subch}$ & 5 \& \mbox{10 \acp{PRB}}\\ 
    Control channel configuration & Adjacent\\
    Retransmissions & Disabled\\
    Keep probability & 0.5 \\
    Min. \& Max. time for the allocation, $T_1$\&$T_2$ & $1$~ms \& $100$~ms \\
    \hline \hline
    \end{tabular}
}
\end{table}

\section{Validation of the Proposed Methodology}

In this section we validate the proposed methodology considering IEEE~802.11p and LTE-V2X sidelink. First, the proper PER value $\beta$ is derived to get a best approximation of the PER vs. SINR curves with the step functions. Then, we estimate the best-fit implementation loss $\hat{\alpha}$ in different environments and in both \ac{LOS} and \ac{NLOS} conditions. Finally, based on the estimated implementation loss, we set the SINR threshold $\gamma_\text{th}$ for the step function and use it to evaluate the performance in terms of \ac{PRR} and \ac{IPG}. The \ac{PRR} is the percentage of the packets correctly received at a given distance, and the \ac{IPG} is the time interval between two consecutive correct receptions at the same receiver from the same transmitter within a given range (here set to 150~m).  Results are obtained using the open-source simulator WiLabV2Xsim \cite{TodBarCamMolBerBaz:21},\footnote{\url{https://github.com/V2Xgithub/WiLabV2Xsim}} with the main simulation parameters listed in Table~\ref{Tab:Settings}.

\subsection{Derivation of the PER Value $\beta$ for the Step Function Approximation} \label{sec:per_value}

\begin{figure}
\centering
\includegraphics[keepaspectratio,width=0.8\columnwidth]{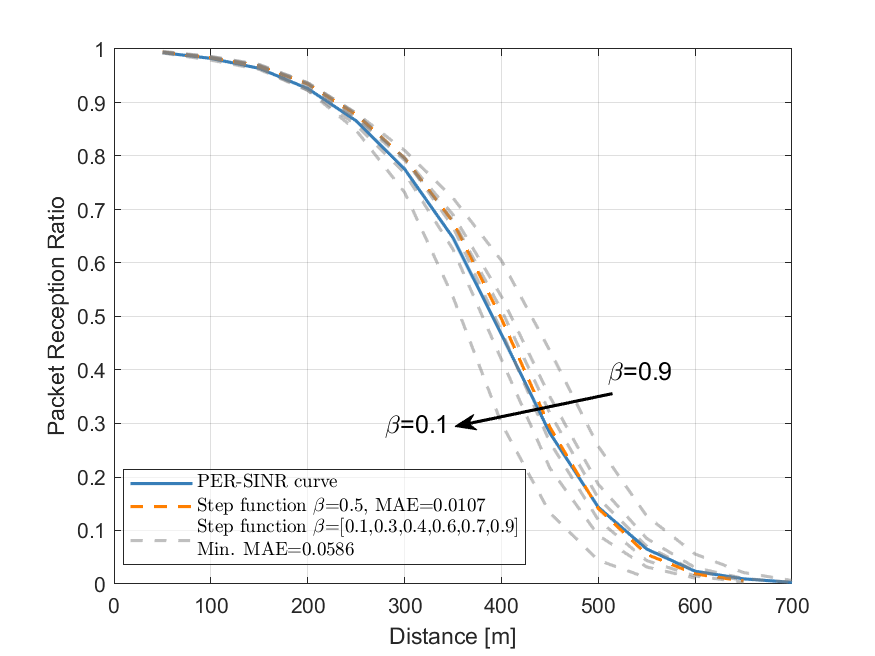}
\caption{PRR vs. distance assuming the PER vs. SINR curve (solid curve) or the step function with different values of $\beta$ (dashed curves). LTE-V2X MCS 7, 350 bytes, 100 vehicles/km. The orange dashed curve corresponds to $\beta=0.5$.}
\label{fig:evaluate_beta}
\end{figure}

The suitable PER value $\beta$ has been derived following the approach described in Section~\ref{sec:approximation}. As an example, Fig.~\ref{fig:evaluate_beta} compares the communication performance in the highway LOS scenario, with 100 vehicles/km, LTE-V2X MCS~7, and 350~bytes packet size. As observable from the curves and confirmed by the minimization of the MAE, the best comparison with the solid line (i.e., the one obtained with the PER vs. SINR curve) is achieved when $\beta=0.5$, which corresponds to the orange dashed curve.

A number of additional results, assuming different technologies, MCSs, packet sizes, and  vehicle densities are also evaluated and reported in Table~\ref{Tab:mae}. Note that $\beta=0.5$ represents the best approximation of the PER vs. SINR model under any settings. It can also be noted that the MAE is always very small, confirming that the step function is a good approximation of the curve when looking at the network level simulations. Given the discussed results, in the rest of the paper $\beta=0.5$ is used.

\begin{table}[!t]
    \caption{Mean absolute error between the performance in terms of PRR vs. distance when comparing the use of the PER vs. SINR curve and the step function. }
    \label{Tab:mae}
    \centering
    \adjustbox{width=\textwidth}{
        \begin{tabular}{c|cccc|cccc}
        \hline \hline
            \multirow{3}{*}{$\beta$} & \multicolumn{4}{c|}{IEEE 802.11p} & \multicolumn{4}{c}{LTE-V2X}\\ \cline{2-9}
            ~ & \multicolumn{2}{c|}{MCS 2, 350 bytes} & \multicolumn{2}{c|}{MCS 4, 550 bytes} & \multicolumn{2}{c|}{MCS 7, 350 bytes} & \multicolumn{2}{c}{MCS 11, 550 bytes} \\ \cline{2-9}
            ~ & \multicolumn{1}{c|}{100 v/km} & \multicolumn{1}{c|}{400 v/km} & \multicolumn{1}{c|}{100 v/km} & \multicolumn{1}{c|}{400 v/km} & \multicolumn{1}{c|}{100 v/km} & \multicolumn{1}{c|}{400 v/km} & \multicolumn{1}{c|}{100 v/km} & \multicolumn{1}{c}{400 v/km} \\ \hline
            0.1 & 0.0621 & 0.0430 & 0.0812 & 0.0702 & 0.0586 & 0.0442 & 0.0668 & 0.0397 \\ 
            0.3 & 0.0184 & 0.0157 & 0.0278 & 0.0254 & 0.0215 & 0.0166 & 0.0237 & 0.0130 \\ 
            0.4 & 0.0086 & 0.0055 & 0.0149 & 0.0130 & 0.0125 & 0.0066 & 0.0144 & 0.0058 \\ 
            0.5 & \textbf{0.0079} & \textbf{0.0039} & \textbf{0.0106} & \textbf{0.0073} & \textbf{0.0107} & \textbf{0.0028} & \textbf{0.0077} & \textbf{0.0050} \\ 
            0.6 & 0.0184 & 0.0147 & 0.0158 & 0.0112 & 0.0129 & 0.0047 & 0.0133 & 0.0099 \\ 
            0.7 & 0.0307 & 0.0194 & 0.0275 & 0.0230 & 0.0212 & 0.0115 & 0.0227 & 0.0189 \\ 
            0.9 & 0.0372 & 0.0257 & 0.0691 & 0.0521 & 0.0470 & 0.0296 & 0.0289 & 0.0334 \\ \hline \hline
        \end{tabular}
    }
\end{table}

\begin{figure}
\centering
\includegraphics[keepaspectratio,width=0.8\columnwidth]{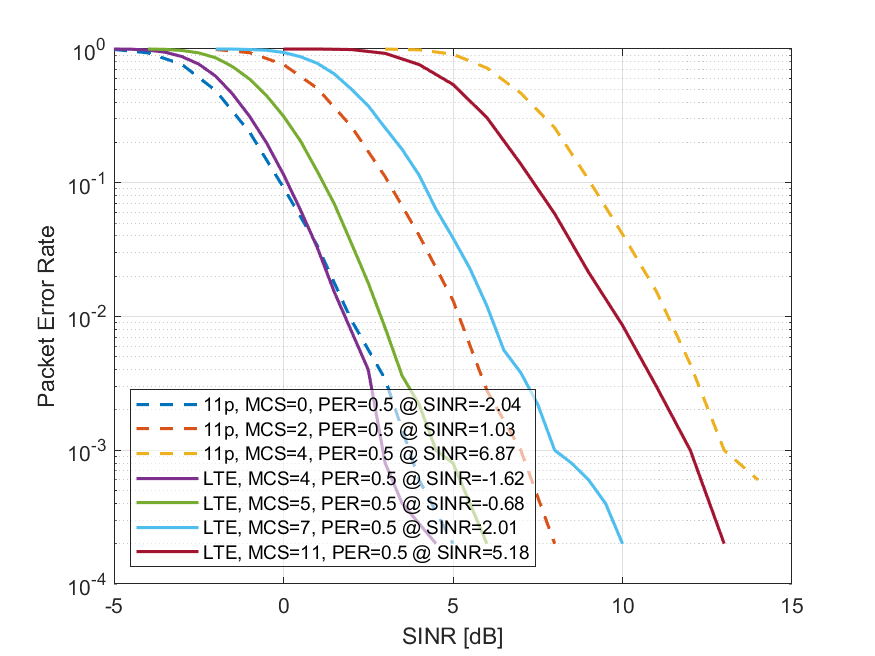}
\caption{PER vs. SINR curves for IEEE 802.11p and LTE-V2X as a function of SINR, in the highway LOS scenario for some of the possible MCSs and assuming packets of 350 bytes.}
\label{fig:PERmeasured}
\end{figure}

\begin{figure}
\centering
\includegraphics[keepaspectratio,width=0.8\columnwidth]{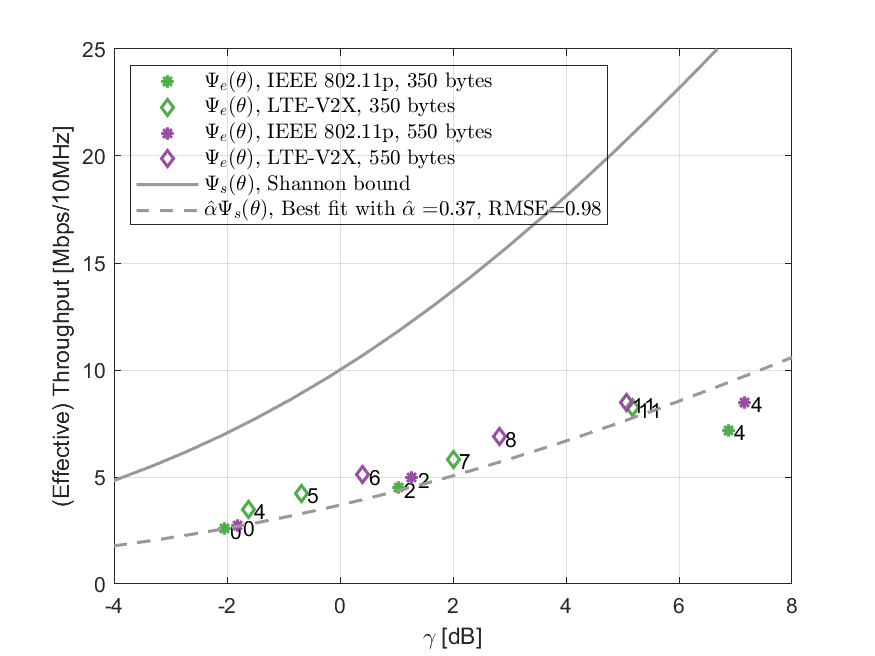}
\caption{Impact of the implementation loss. The colored symbols show the effective throughput $\Psi_\text{e}(\boldsymbol \theta)$ vs. SINR threshold $\gamma_\text{th}(\boldsymbol \theta)$ for the system settings for which the PER vs. SINR curve is available, with the numbers next to them representing the MCS indexes. The solid curve is the Shannon bound corresponding to the SINR value. The dashed line is the best fit curve with the implementation loss $\hat{\alpha}$.
}
\label{fig:ThroughputVsSinr}
\end{figure}

\subsection{Implementation Loss $\hat{\alpha}$ in the Considered Scenarios}
\label{sec:estimatedalpha}

Based on the dataset $\{\boldsymbol{\theta}_i | i=1,2,\ldots,N \}$ of measured PER vs. SINR curves (part of them are presented in Fig.~\ref{fig:PERmeasured}) for both the IEEE 802.11p and LTE-V2X technologies,  $\gamma_\text{th}(\boldsymbol \theta_i)$ are obtained with $\beta=0.5$. Then, the estimated $\hat{\alpha}$ is derived from \eqref{eq:estimatedalpha}. Fig.~\ref{fig:ThroughputVsSinr} represents the result of this operation 
for the highway LOS scenario, by showing the effective throughput varying the SINR threshold. In particular, the continuous curve corresponds to the Shannon bound. Then, each symbol indicates the effective throughput and the corresponding SINR for one of the settings for which the PER vs. SINR is available; the packet size is indicated by the color, the technology by the symbol shape, and the MCS index by the number written near to the symbol. The dashed curve shows the curve obtained using the optimized implementation loss, which is in this case equal to 0.37. 
The figure confirms that the model resulting from the proposed methodology 
with the estimated $\hat{\alpha}$ well approximates multiple system settings. Results corresponding to other scenarios are reported in Table~\ref{Tab:lossFactor}.

\begin{table}
\caption{Implementation loss $\hat{\alpha}$ in different scenarios, considering different configurations $\{\boldsymbol \theta_i | i=1,2,\ldots,N \}$, varying the MCS index and packet size. The RMSE of the effective throughput with respect to the approximation is reported.
\label{Tab:lossFactor}}
\centering
\begin{tabular*}{\textwidth}{m{2.5cm}m{0.6cm}<{\centering}m{3.5cm}m{1.5cm}<{\centering}m{0.6cm}<{\centering}m{2.5cm}<{\centering}}
\hline \hline
Scenarios & $N$ & MCS (802.11p) \& (LTE) & $P_\text{b}$ [bytes]& $\hat{\alpha}$ & RMSE [Mb/s]\\
\hline
Crossing NLOS & 7 & (0, 2, 4) \& (4,5,7,11)  & 350 & 0.25 & 0.82\\
Highway LOS & 13 & (0, 2, 4) \& (4~8,11) & 350, 550 & 0.37 & 0.98\\
Highway NLOS & 13 & (0, 2, 4) \& (4~8, 11) & 350, 550 & 0.24 & 0.80\\
Urban LOS & 7 & (0, 2, 4) \& (4,5,7,11) & 350, 550 & 0.32 & 0.99\\
\hline \hline
\end{tabular*}
\end{table}

\subsection{Validating Network Level Results}
We now assess the effectiveness of the proposed physical layer abstraction by evaluating the \ac{V2X} communication performance both adopting the PER vs. SINR curves (as in Fig.~\ref{fig:PERmeasured}) and the model deriving from proposed methodology (with the threshold SINR $\hat{\gamma}_\text{th}$ obtained through \eqref{eq:get_threshold}). Please note that the settings adopted are necessarily among those for which the curve is available, whereas the proposed methodology would allow us to also consider the other settings. 


\begin{figure}[!htbp]
\centering
\begin{subfigure}{1\textwidth}
  \centering
  \includegraphics[keepaspectratio, width=0.8\columnwidth]{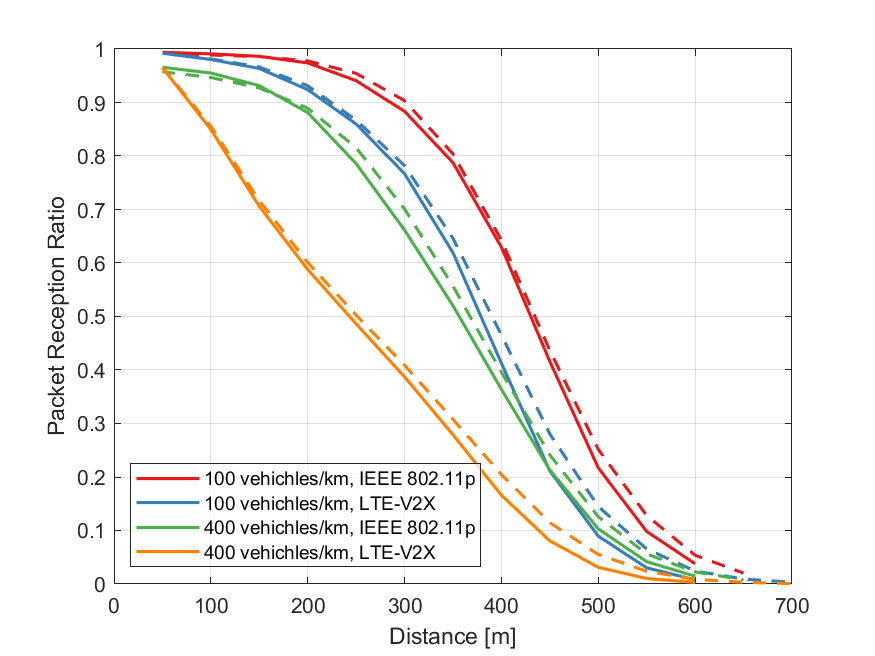}
  \caption{\centering PRR vs. distance.}
  \label{fig:PRRall}
\end{subfigure}\hfill\vskip 0.3cm
\begin{subfigure}{1\textwidth}
  \centering
  \includegraphics[keepaspectratio, width=0.8\columnwidth]{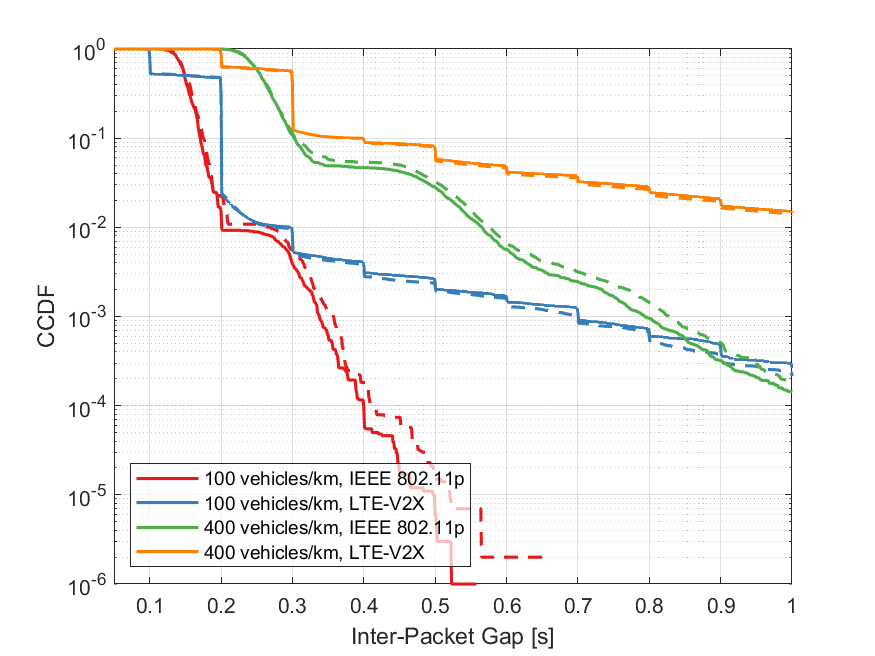}
  \caption{\centering CCDF of IPG.}
  \label{fig:IPGall}
\end{subfigure}
\caption{PRR vs. distance and CCDF of IPG obtained using the proposed step function approximation (solid curves) and the PER vs. SINR curves of Fig.~\ref{fig:PERmeasured} (dashed curves). Results are obtained in a highway scenario with LOS conditions.}
\label{fig:All}
\end{figure}

Fig.~\ref{fig:PRRall} shows the \ac{PRR} varying the transmission distance and with a density of 100 or 400 vehicles/km. As illustrated in the figure, the results based on the proposed methodology are very close to those with the PER vs. SINR curve. The difference between the evaluated performance increases slightly for larger values of the transmission distance. Similar results shown in Fig.~\ref{fig:IPGall}, which plots the \ac{CCDF} of the \ac{IPG}, demonstrate that the proposed methodology can also evaluate \ac{IPG} with high accuracy. Overall, the slightly increased error, in terms of PRR when the distance gets larger (in Fig.~\ref{fig:PRRall}), or in terms of \ac{IPG} when a longer value is observed (in Fig.~\ref{fig:IPGall}), appears negligible. 




Please remark that adopting the proposed methodology, curves similar to those in Fig.~\ref{fig:All} can be easily obtained for both IEEE~802.11p and  LTE-V2X sidelink, in any scenario of Table~\ref{Tab:lossFactor}, for any packet size and for any \ac{MCS}. Differently, if the more accurate reference model was used, new curves derived from additional measurements or link-level simulations were required for most of the possible configurations.

\section{Conclusion}
 In this paper, a new methodology has been proposed for modelling the physical layer in the network-level evaluation of direct V2X communications technologies. 
 The proposed methodology is general and is a low-complexity, accurate alternative to the use of PER vs. SINR curves, which are normally available only for a few configurations. The resulting model is characterized by a single parameter called implementation loss, here calculated for various scenarios. 
 
The proposed approach has been validated based on network-level simulations with both IEEE~802.11p and LTE-V2X sidelink technologies, while benchmarking results are obtained using the PER vs. SINR curves.
Results show that the proposed methodology provides an accurate assessment of \ac{V2X} communications without requiring costly measurement campaigns or computationally intensive simulations at the link level. 
 
 
In future works, the plan is to adapt and validate the model to new technologies, including IEEE 802.11bd and NR-V2X sidelink. Nevertheless, as the proposed approach demonstrated high generality when used for IEEE~802.11p and LTE-V2X sidelink, despite their strong differences in the PHY and MAC protocols, it is likely that the same model is still applicable for other technologies. 

\acknowledgments{We would like to thank the China Scholarship Council that is supporting Wu Zhuofei during his visiting scholarship at the University of Bologna.}





\begin{adjustwidth}{-\extralength}{0cm}

\reftitle{References}


\bibliography{myBib}

\end{adjustwidth}
\end{document}